\documentclass[aps,shownopacs,superscriptaddress,nofootinbib,preprint,11pt]{revtex4}
\usepackage{hyperref}
\usepackage{amsmath}
\usepackage{array}
\newcolumntype{L}[1]{>{\raggedright\let\newline\\\arraybackslash\hspace{0pt}}m{#1}}
\newcolumntype{C}[1]{>{\centering\let\newline\\\arraybackslash\hspace{0pt}}m{#1}}
\newcolumntype{R}[1]{>{\raggedleft\let\newline\\\arraybackslash\hspace{0pt}}m{#1}}
\usepackage{float}
\usepackage{graphicx}
\usepackage{epsfig}
\usepackage{psfrag}
\usepackage{color}
\usepackage{slashed}

\usepackage{amsfonts}
\usepackage{amssymb}
\usepackage{tikz}
\usepackage{tikz}
\usetikzlibrary{positioning,arrows}
\usetikzlibrary{decorations.pathmorphing}
\usetikzlibrary{decorations.markings}

\newcommand*{\be}{\begin{equation}}
\newcommand*{\ee}{\end{equation}}
\newcommand*{\bea}{\begin{eqnarray}}
\newcommand*{\eea}{\end{eqnarray}}

\newcommand{\comment}[1]{}

\newcommand{\cref}[1]{Chapter~\ref{c.#1}}

\def\beq{\begin{equation}}
\def\eeq{\end{equation}}
\def\bea{\begin{eqnarray}}
\def\eea{\end{eqnarray}}
\def\ba{\begin{array}}
\def\ea{\end{array}}
\def\bi{\begin{itemize}}
\def\ei{\end{itemize}}
\def\be{\begin{enumerate}}
\def\ee{\end{enumerate}}
\def\bc{\begin{center}}
\def\ec{\end{center}}
\def\bt{\begin{table}}
\def\et{\end{table}}
\def\btb{\begin{tabular}}
\def\etb{\end{tabular}}

\def\t{\tilde}

\def\lsim{\raise0.3ex\hbox{$\;<$\kern-0.75em\raise-1.1ex\hbox{$\sim\;$}}}
\def\gsim{\raise0.3ex\hbox{$\;>$\kern-0.75em\raise-1.1ex\hbox{$\sim\;$}}}
	
 \preprint{
 	TIFR/TH/16-31
 }
\begin{document}

\title{Dissecting new physics models through kinematic edges}
\author{Abhishek M. Iyer}
\email{abhishek@theory.tifr.res.in}
\address{Department of Theoretical Physics, Tata Institute of Fundamental Research, Homi Bhabha Road, Colaba, Mumbai 400 005, India}

\author{Ushoshi Maitra}
\email{ushoshi@theory.tifr.res.in}
\address{Department of Theoretical Physics, Tata Institute of Fundamental Research, Homi Bhabha Road, Colaba, Mumbai 400 005, India}

\begin{abstract}
Kinematic edges in the invariant mass distributions of different final state particles are typically a signal of new physics.
In this work we propose a scenario wherein these edges could be utilised in discriminating between different classes of models. To this effect, we consider the resonant production of
a heavy Higgs like resonance ($H_1$) as a case study. Such states are a characteristic feature of many new physics scenarios beyond the Standard Model (SM). In the event of a discovery, it is essential to identify the true nature of the underlying theory. In this work we propose a channel, $H_1 \rightarrow t_2t$, where $t_2$ is a vector-like gauge singlet top-partner that decays into $Wb,~Zt,~ht$.  Invariant mass distributions constructed out of these final states are characterized by the presence of kinematic edges, which are unique to the topology under consideration. Further, since all the final state particles are SM states, the position in the edges of these invariant mass distributions can be used to exclusively determine the masses of the resonances. Observation of these features are meant to serve as a trigger, thereby mandating a more detailed analysis in a particular direction of parameter space. The absence of these edge like features, in the specific invariant mass distributions considered here, in minimal versions of supersymmetric models (MSSM) also serves as a harbinger of such non MSSM-like scenarios.

\end{abstract}
\vskip .5 true cm

%\pacs{73.21.Hb, 73.21.La, 73.50.Bk}
\maketitle
\section{INTRODUCTION}
The presence of an elementary scalar in the Standard Model (SM) provides the most compelling reason to expect new physics at the  TeV scale and beyond. 
While new physics candidates may differ quite significantly with respect to the underlying theory, 
there could be similarities in the properties of the new particles predicted by these scenarios.
 For instance several models are characterized by the presence of a heavy scalar with Higgs like couplings. 
One class of models (say Class A) which are characterized by these heavy scalars include two Higgs doublet model, supersymmetric model \textit{etc.}. 
Alternatively, these scalars can also arise as the Kaluza-Klein (KK) excitations of the bulk Higgs in the extra-dimensional models or supersymmetric extensions of Little Higgs scenarios~\cite{Roy:2005hg, Csaki:2008se}. 
We refer to these kind of models as Class B. In the event of a discovery of such excited scalars, it is essential to identify the true nature of physics beyond the SM.
In this paper we propose an analysis that help us to recognize such distinct features.\\ 

The models are segregated into two classes: A and B introduced earlier.
The latter is characterized by the presence of additional vector-like fermions. 
For instance, the vector-like fermions could correspond to the KK excitations of the fermions 
in the extra dimensional models~\cite{Agashe:2003zs}
with the lightest one generally corresponding to the top-partner or they can also arise in the extended
Little Higgs model~\cite{Roy:2005hg, Csaki:2008se}.
% \slashed{ excitations are absent in the minimal versions of models belonging to Class A.} 
Using this difference, we attempt to devise a unique signature which is a trademark of models belonging to Class B.\\

The phenomenology of the heavy top partners at the LHC has been discussed in details in Ref.~\cite{Gopalakrishna:2013hua, Gopalakrishna:2011ef,DeSimone:2012fs, Contino:2006qr,Vignaroli:2012sf, Anastasiou:2009rv, Kribs:2010ii,
Carena:2006jx, Han:2003wu, Matsumoto:2008fq, Buchkremer:2013bha, Banfi:2013yoa, Li:2013xba, Gripaios:2014pqa,Chala:2014mma, Endo:2014bsa, Dolan:2016eki}.
Similarly the phenomenology of a heavy scalar coming from a general Higgs sector or a warped extra dimension at the LHC has been discussed in Ref.~\cite{Dumont:2014wha, Mahmoudi:2016aib}.
Recently, ATLAS and CMS have searched for a heavy Higgs-like resonance in the $WW^{*}$,
diphoton, $ZZ$ and $hh$ channels~\cite{Aad:2015kna, Aad:2014ioa, ATLAS-CONF-2016-059, ATLAS-CONF-2016-079, ATLAS-CONF-2016-056, Sirunyan:2016cao}.
Although, translating the maximum observed cross section
as an exclusion on the mass of the heavy Higgs is highly
model dependent,  one can safely assume 
 that a heavy scalar beyond 1 TeV is still compatible with these search limits. These searches
for heavy Higgs as well as heavy top partners are in general carried out independent of each other. Thus, even if there
is a discovery in any of these search modes, it is difficult to pin-point the right class of model.

The aim 
of this paper is to present a unified search strategy for the heavy Higgs scalar and the heavy top-partner. This would eventually serve as a litmus test in distinguishing
models belonging to Class-A  from Class-B. \footnote{It is relevant to note at this point that  the heavy scalar can be replaced by any other spin particle with similar mass, which couples to the vector like top. The analysis presented in the work proceeds in an exactly similar fashion. A detailed discussion to this effect is given in Section \ref{results} }\\ 
% The bounds coming from precision electroweak observables \cite{Ellis:2014dza} and LHC direct searches for a heavy top-like quarks rule out any heavy quark below 600 GeV~\cite{Aad:2015kqa, Aguilar-Saavedra:2013qpa, Ellis:2014dza}. 

%Recently,~\cite{Chala:2014mma} has studied the phenomenology of a spin-1 resonance decaying to heavy top partners. 
%We propose
%that such channels are powerful in discriminating models belonging to Class-A from Class-B.

\section{Model} Consider a simplified model with a heavy Higgs-like scalar ($H_{1}$) and a vector-like gauge singlet fermion ($t'$).
The relevant couplings of the scalar $H_{1}$ to $t'$, gluons\footnote{$H_1$ is assumed to be produced through gluon fusion diagram with SM top quarks propagating in the loop} and third generation quarks are governed by the following effective Lagrangian:
\begin{equation}
\mathcal{L}_{NP}\supset G^a_{\mu\nu}G^a_{\mu\nu}H_{1}+\left(Y_t\bar  Q_3H_1t'+Y_t\bar Q_3H_{1} t + Y_t\bar Q_3H  t'+M_{t'}\bar t't' + Y_t\bar Q_3H t+h.c\right)
\label{effective_lagrangian}
\end{equation} 
 where $G^a_{\mu\nu}$ is the field strength tensor for the gluons. We have assumed the rest of the vector-like spectrum to be heavy and is decoupled from the effective low energy theory.
 Without  loss of generality, we assume the coupling strength of the scalar $H_{1}$ to $Q_3$ and $t'$ to be the same as $Y_t$, the top Yukawa coupling \footnote{The size of this coupling comes into play when considering the branching fraction of $H\rightarrow t_2~t$. In this scenario we assumed a branching fraction of 50\% for this model. For a warped model the branching fraction is dominated in $H1\rightarrow t t$ mode with the gauge bosons mode suppressed due to orthonormality. The Branching fraction of $H1\rightarrow t t$ can be adjusted by playing with the localization parameter  of the bulk scalar. Observation of these edges requires the accumulation of a certain minimum number of signal points. Lower branching fractions would suffer with larger luminosity reaches, primarily due to the lower production cross section of a heavy scalar.}.
 Since the heavy scalar $H_{1}$ has Higgs-like interaction, its decay to a pair of $t\bar{t},WW,ZZ,hh$ is common for both classes of models under consideration.
As mentioned earlier, models belonging to Class B are characterized by the presence of an additional vector-like states.
% In the minimal setup, the interaction Lagrangian for a top partner  $t'$ is given as
%\begin{equation}
%\mathcal{L}_{t'th}\supset Y_t\bar Q_3H  t'+M_{t'}\bar t't' + Y_t\bar Q_3H t
%\label{tprime}
%\end{equation}
 The fourth  term in the parenthesis in Eq. \ref{effective_lagrangian} induces a mass-mixing between the SM top and its vector-like counterpart. 
 The mass matrix, in the basis $(t,t')$, parametrizing  this mixing is  given by:
 \begin{equation}
 \mathcal{M}_{tt'} =  \begin{bmatrix}
 	\frac{Y_{t} v}{\sqrt{2}}&& \frac{Y_{t} v}{\sqrt{2}}\\
 	 0 && M_{t'}
 	\end{bmatrix}
 \end{equation} where $v$ represents vacuum expectation value (vev) of the Higgs.

 In the presence of this mixing the mass-basis is related to the interaction basis as:
 \begin{equation}
 \begin{bmatrix}
 	t_1\\t_2
 	\end{bmatrix}_{L,R}=\mathcal{O}_{2\times 2}\begin{bmatrix}
 	t\\t'
 	\end{bmatrix}_{L,R}
 \end{equation}
where $t_1$ is now identified as the SM top and $t_2$ is the heavier partner. $\mathcal{O}$ is the $2\times 2$  rotation matrix to move from the interaction basis to the mass basis.
Henceforth, the top will be denoted as $t$ for convenience. An example of a complete model in this case would be a warped extra dimensional model \cite{Randall:1999ee,Gherghetta:2000qt}. 
The masses of the heavy partner of the $SU(2)$ singlet is light as the corresponding bulk field is localized closer to the IR brane (The bulk localization parameter is $c\sim0$).
The doublets do not enjoy a similar localization as due to constraints from $Zb\bar{b}$. As a result, the corresponding $KK$ partners are heavier.
The masses of $n=1$  KK partners of $W$ and $Z$ are significantly heavier due to constraints from precision electroweak and flavour physics.
On the other hand, the mass of the $n=1$ KK partner of Higgs is not as severely constrained and can be as low as 1 TeV.

\begin{figure}[!t]
	\begin{center}
		\begin{tabular}{cc} 
			\includegraphics[width=8cm]{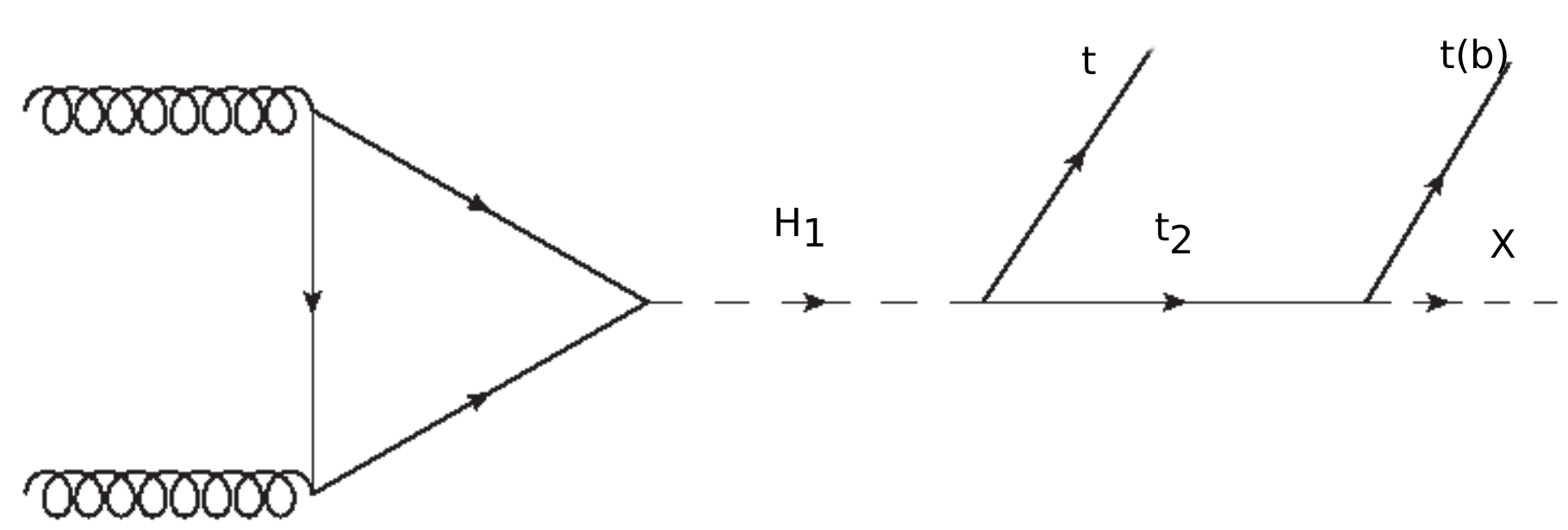}
% 			&\includegraphics[width=8.2cm]{edge2.png}
		\end{tabular}
	\end{center}
	\caption{\it{ggF production of $H_{1}$ and its decay to $t~t (b)~X$}}
	\label{tbw}
\end{figure}

The presence of an additional  fermion $t_2$ opens up an additional channel for $H_{1}$ to decay i.e  $H_1\rightarrow tt_2$. For the setup under consideration, $t_2$ can only decay into $t~(b)~+~X$ where $X=W,h,Z$. 
The branching fraction of $t_2$ decaying to the gauge bosons is governed by its interaction to the scalar degrees of the Higgs doublet $H$. Two out of four degrees of freedom correspond to the longitudinal polarization of $W^{\pm}$, 
while one is that of the $Z$-boson. The remaining is the SM Higgs boson $h$. Consequently,  one can roughly estimate the branching rates to be
\begin{equation}
B.R(t_2\rightarrow b+W)\sim 50\%;\;\;B.R(t_2\rightarrow t+h)\sim 25\%;\;\;B.R(t_2\rightarrow t+Z)\sim 25\%
\end{equation}
All the channels corresponding to $H_{1}~\rightarrow~tt_2~\rightarrow~tt(b)X$ as depicted in Fig.~\ref{tbw} are characterized by distinct kinematic endpoints in certain invariant mass distributions. This unique feature not only distinguishes it from SM backgrounds, but also from models belonging to Class A, serving as a smoking gun signal for Class B.

Kinematic variables like $M_{T_2}$ have been used in different SUSY searches \cite{Lester:2001zx,Lester:1999tx,Allanach:2000kt,Meade:2006dw,Cheng:2007xv}.
 We would like to emphasize that our current analysis is to use the known technique of kinematic edges  as a smoking gun towards the presence of certain specific models. To this effect, we have constructed a topology with a heavy Higgs and a vector like top partner. This leads to final states with for instance top, bottom and W (for the leading decay mode of $t_2$) which are visible with known masses. As discussed, the 
invariant mass distribution of top and bottom and bottom and lepton (from W) exhibit edges in the kinematic endpoints. This `edgy' feature in this particular final state is only a characteristic of models which have a vector like top partner.
For a cascade decay having $P_{1}~\rightarrow~P_{2} d_{1}~\rightarrow~d_{2}d_{3}$,  upper edge in the invariant mass distribution involving final state particles $d_{1}, d_{2}$ is given by ~\cite{Lester:2001zx,Lester:2006yw}
\begin{eqnarray}
	m_{edge}^{2} = m_{d_1}^{2} + m_{d_2}^{2}  + \frac{f(m_{P_1}, m_{P_2}, m_{d_1}) f (m_{P_2}, m_{d_2}, m_{d_3})-g(m_{P_2}, m_{d_1}, m_{P_1}) g(m_{P_2}, m_{d_2}, m_{d_3})}{2 m_{P_2}^{2}} 
	%  \nonumber\\
	%                +& \frac{\sqrt{((m_{p_1}^{2} - m_{p_2})^{2} - m_{d_1}^{2})((m_{p_1}^{2} + m_{p_2})^{2} - m_{d_1}^{2})}\sqrt{((m_{p_1}^{2} - m_{p_2})^{2} - m_{d_1}^{2})((m_{p_1}^{2} + m_{p_2})^{2} - m_{d_1}^{2})}}{2 m_{p_2}^{2}} \nonumber \\
	%                + & \frac{(m_{P_1}^{2} - m_{d_1}^{2} - m_{P_2}^2)(m_{P_2}^{2} + m_{d_2}^{2} - m_{d_3}^{2})}{2 m_{p_2}^{2}}
\label{sample}
\end{eqnarray}
where $f(a, b, c)~=~\sqrt{(a^2 - b^2 - c^2)(a^2 + b^2 - c^2)(a^2 - b^2 + c^2)(a^2 + b^2 + c^2)}$ and \\
$g(a, b, c)~=~a^{2} - b^{2} - c^{2}$.\\
Models like MSSM which can also lead to similar final states, do not however exhibit these edges as the final states  are uncorrelated and similar invariant masses in such a case will lead to gradually falling pattern. The Heavy Higgs with a vector like top in this case is only a toy model. The Higgs partner can be replaced to include a $Z'$, Graviton and the analysis proceeds similarly.

We now study each of these channels and define the invariant mass distributions where the kinematic endpoints can be observed.

\subsection{Channel 1 : $H_1\rightarrow tt_2 \rightarrow t b W$}
 This is the most dominant channel where $H_1$  decays to $t$ and $t_2$ with $t_2$ further decaying to $b$ -quark and $W$-boson.
%  It is depicted in Fig. \ref{tbw}.
While the top decays hadronically, we consider the leptonic decay mode $W$- boson. 
 The topology is endowed with the following features:
\begin{itemize}
\item \textbf{Kinematic edge in the  $m_{bt}$ distribution:} 

As discussed the invariant mass distribution of the top and bottom
quarks system is characterized by a kinematic endpoint. The invariant mass of top and bottom system is given by
\begin{equation}
\left(m^{edge}_{tb}\right)^2~=~m_t^2~+~m_{b}^{2}~+~2\left(E_{t}E_{b} -  \textbf{{P}}_{t}.\textbf{{P}}_{b} \right),
\label{edgetb}
\end{equation}
where $t$ is the top quark originating from the heavy Higgs while the $b$-quark emerges from the decay of $t_2$.
The magnitude of the quarks momentum in the rest frame of $t_2$ are given as
\begin{eqnarray}
p_{t}^2=\frac{m_t^4+m_{t_2}^4+m_{H_1}^4-2\left(m_t^2m_{t_2}^2+m_t^2m_{H_1}^2+m_{t_2}^2m_{H_1}^2\right)}{4m_{t_2}^2}\nonumber\\
p_{b}^2=\frac{m_{t_2}^4+m_{b}^4+m_{W}^4-2\left(m_b^2m_{t_2}^2+m_b^2m_W^2+m_{t_2}^2m_{W}^2\right)}{4m_{t_2}^2}
\end{eqnarray}
and $E_{i}^2=m_i^2+p_{i}^2$.
The invariant mass acquires its maximum value when the angle between the top quark and the bottom quark is  $\pi$.
The edge of the invariant mass is a function of the mass of $H_1$ and $t_2$ as shown in Eq.~\ref{sample}.
Right panel of Fig.~\ref{edges} gives the position of the edge in the $m_{tb}$ distribution as a function of $m_{t_2}$. 
It is plotted for two different masses of the heavy scalar $H_{1}$.
 The green curve represents the edge for $m_{H_1}=1100$ GeV and the blue curve is for $1200$ GeV. 
 
 \begin{figure}[!t]
	\begin{center}
		\begin{tabular}{cc} 
 			\includegraphics[width=9cm]{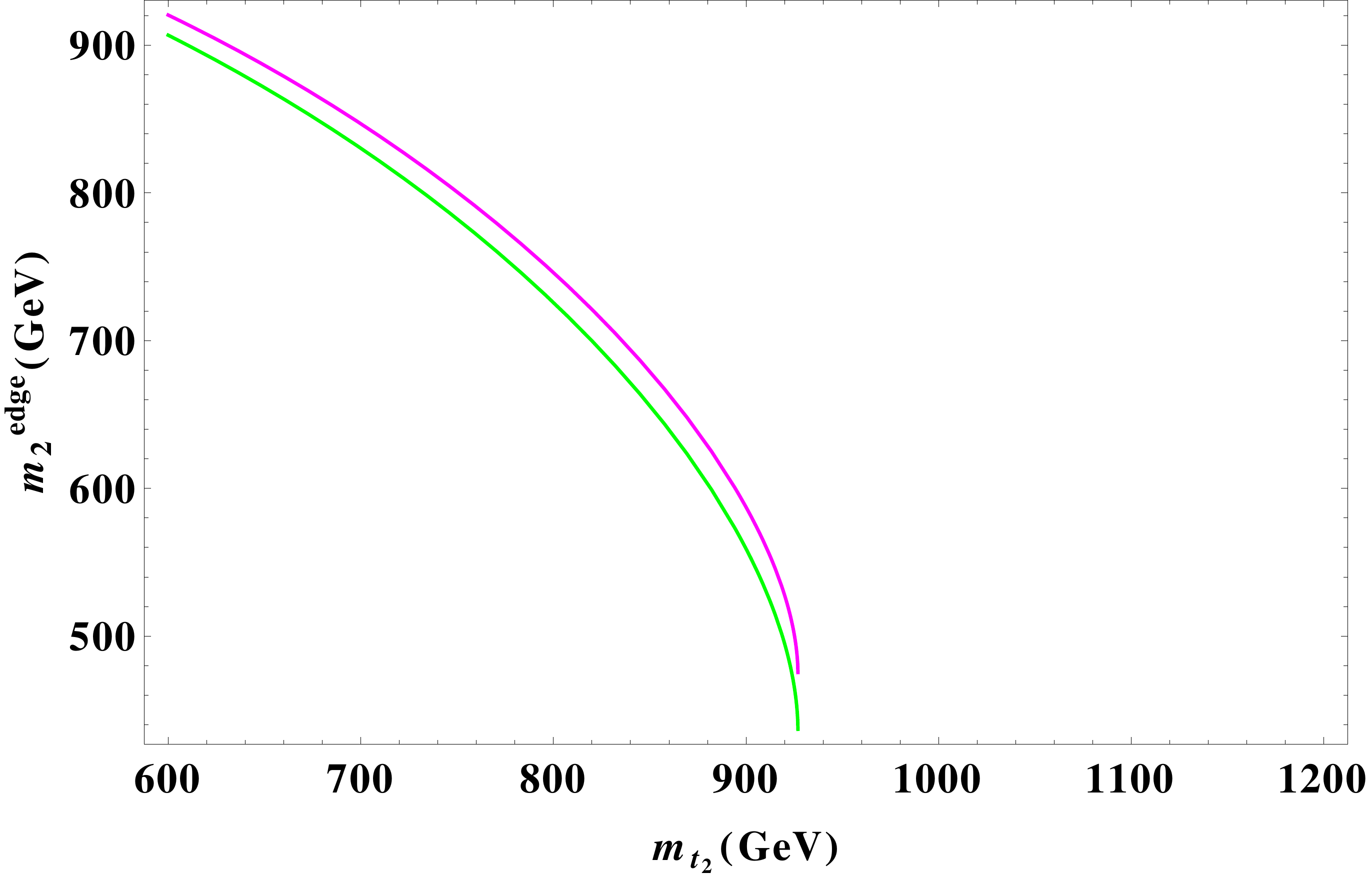}
			\includegraphics[width=9cm]{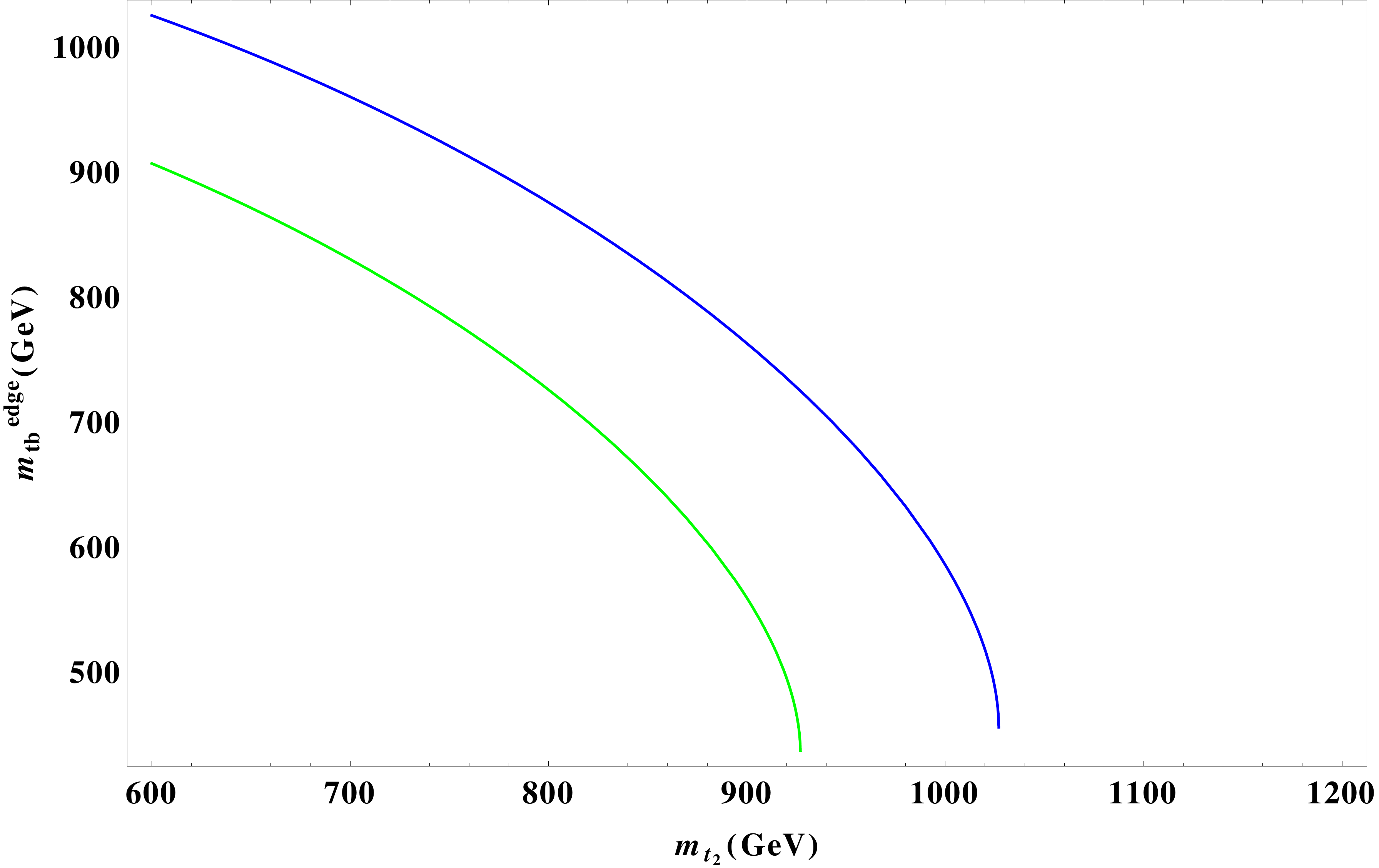}
% 			&\includegraphics[width=8.2cm]{edge2.png}
		\end{tabular}
	\end{center}
	\caption{\it{Variation of the edge of invariant mass for two final state quarks with mass of $t_{2}$ (left figure)}. The green plot represents the edge for $H_1 \rightarrow t b W$ and magenta plot is for $H_{1} \rightarrow t t h.$
	       Variation of the edge of $m_{tb}$ with mass of $t_{2}$ (right figure) for $H_1$ having mass of 1.1 TeV (green) and 1.2 TeV (blue).}
	\label{edges}
\end{figure}
It is clear from Fig.~\ref{edges} that the position of the edge is unique to the choices of masses under consideration.
At this stage it is important to note that we are limited in our choices for the masses for these two particles.
Due to s-channel suppression, the production cross section of the heavy Higgs falls rapidly with increase in mass.
% Increasing the mass of the heavy Higgs $(m_{H_1})$ leads to rapid fall in its production cross-sections. 
Reducing it will necessitate reducing the mass of vector-like quark $(m_{t_2})$ putting it in tension with the
searches for third generation vector-like quarks. Thus, we consider the following three benchmark points: 
\begin{eqnarray}
\text{BP1}:~m_{H_1}~&=&~1.2~\text{TeV},~m_{t_2}~=~600~\text{GeV};\nonumber\\ \text{BP2}:~m_{H_1}~&=&~1.1~\text{TeV},~ m_{t_2}~=~700~\text{GeV};\nonumber\\
\text{BP3}:~m_{H_1}~&=&~1.5~\text{TeV},~ m_{t_2}~=~1000~\text{GeV}
\label{benchmark}
\end{eqnarray}
Due to lower cross section for the heavy Higgs-like scalar we will consider only the dominant decay mode of $t_{2}$ i.e $b$ and $W$ for BP3.
The parton-level distribution for $m_{tb}$ is given in the left panel of Fig.~\ref{channel1_parton}.
Clearly, the distribution exhibits a kinematic edge for both the benchmark points in Eq. \ref{benchmark}. 

\begin{figure}[!t]
	\begin{center}
		\begin{tabular}{cc} 
			\includegraphics[width=9cm]{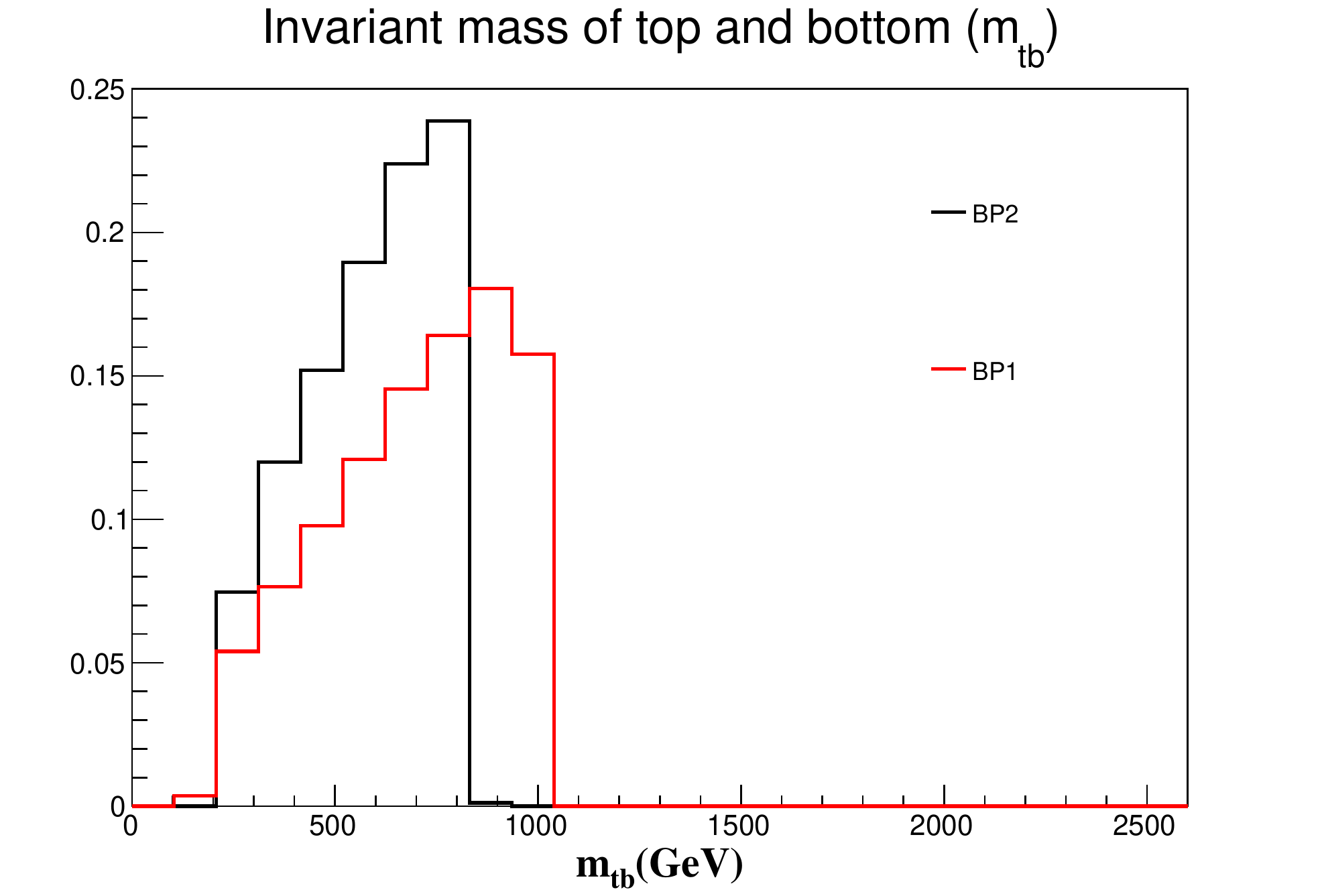}
			\includegraphics[width=9cm]{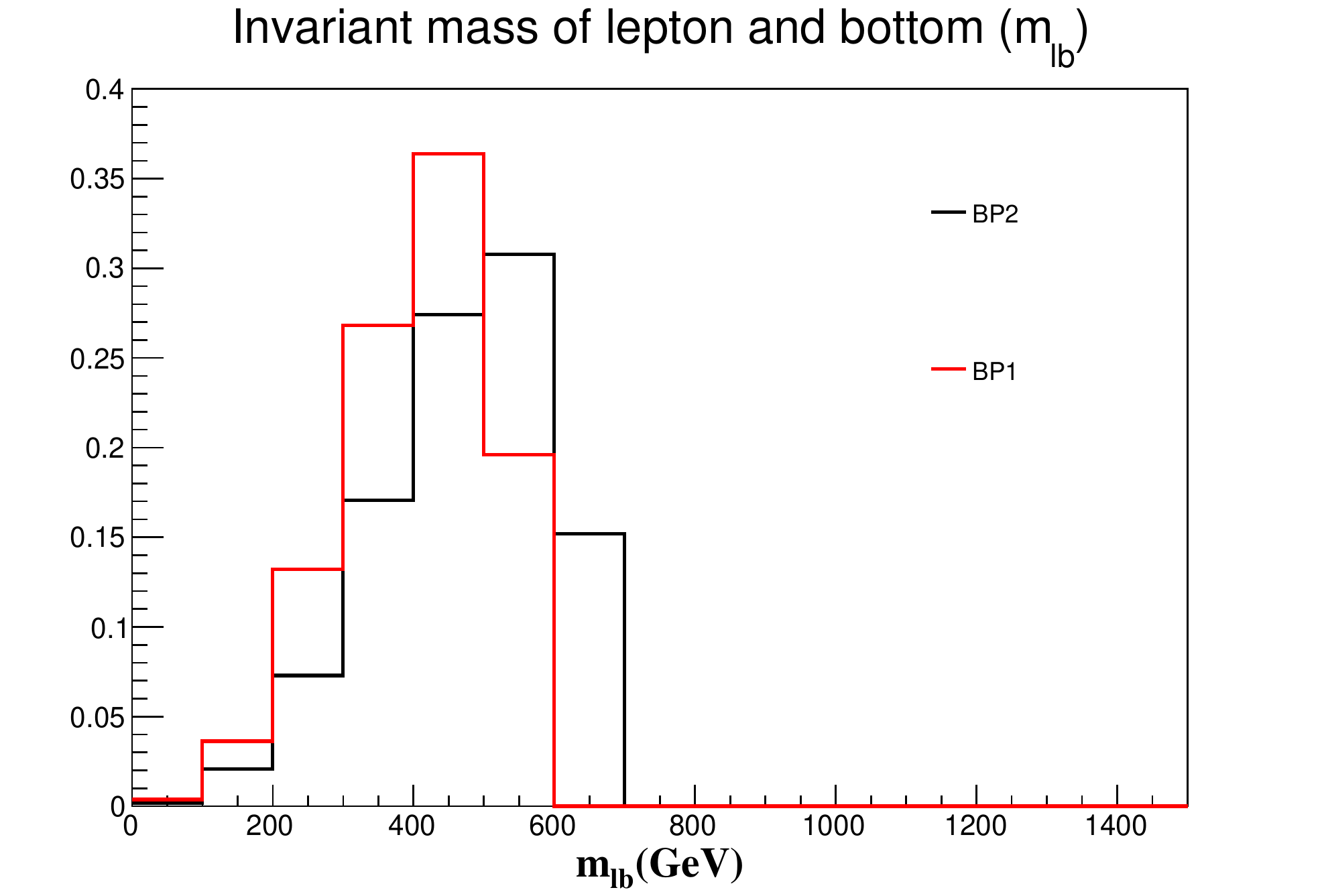}
			% 			&\includegraphics[width=8.2cm]{edge2.png}
		\end{tabular}
	\end{center}
	\caption{\it{Parton-level distributions for $m_{tb}$ (left) and $m_{lb}$ (right) for the two benchmark points, BP1 (red) and BP2 (black)}}
	\label{channel1_parton}
\end{figure}
Right panel of Fig.~\ref{channel1_parton} shows the corresponding distribution for $m_{bl}.$

\item \textbf{ Kinematic edge in the $m_{lb}$ distribution:} In addition to $m_{tb}$, the invariant mass distribution  of the bottom quark and the lepton also
exhibits distinct edge given by
\begin{equation}
\left(m^{edge}_{bl}\right)^2~=~m_b^2~+~2\left(E_{b}E_{l} +  |\textbf{P}_{b}||\textbf{P}_{l}| \right)
\label{edgebl}
\end{equation}
where the lepton ($l$) originates from the $W$ decay.
The magnitude of the quarks momentum in the rest frame of $W$ are given as
\begin{equation}
p_{l}^2=\frac{m_W^2}{4}~,~p_{b}^2=\frac{m_{t_2}^4+m_{b}^4+m_{W}^4-2\left(m_b^2m_{t_2}^2+m_b^2m_W^2+m_{t_2}^2m_{W}^2\right)}{4m_{W}^2}
\end{equation}
and $E_{i}^2=m_i^2+p_{i}^2$. 
% Fig \ref{} shows the parton-level distribution for $m_{tl}$ for $m_H=1.2$ TeV and $m_{t_2}=600$ GeV.
\item \textbf{ Kinematic edge in $m_{tbl}$ distribution:} For completeness, we also note that the invariant mass of top, bottom and lepton system also shows behaviour similar to $m_{tb}$ and $m_{bl}$. The distribution has an edge
at mass of the heavy resonance i.e. $m_{H_1}$. The distribution does not reveal any further information which can add to ones obtained from $m_{tb}$ and $m_{bl}$ and hence, will not be considered further.
\end{itemize}
The parton-level plots in Fig \ref{channel1_parton} are generated by implementing the Lagrangian given in Eq. \ref{effective_lagrangian}  in 
FEYNRULES~\cite{Christensen:2008py} and interfacing it with MADGRAPH~\cite{Alwall:2014hca}.
For a given benchmark point, say BP2 , substitution of the masses in Eq. \ref{edgetb} gives  $m^{edge}_{tb}$ at 840 GeV which
is roughly the location of the edge in the plot. Similar conclusion hold for the other benchmark points.
% From the position of the kinematic endpoint in the $\left(m^{edge}_{bt}\right)$ distribution, and the known masses of the quarks and W-boson, one can determine 
% the mass of the heavy Higgs in terms of the mass of the vector-like quarks using Fig \ref{edges}. 
Given the fact that we are restricted in our choice of the masses for the heavy resonances, Fig \ref{edges} can be used to determine the masses of $H_{1}$ and $t_{2}$ exclusively.
The mass of $t_2$ determined from Fig \ref{edges}  can be validated by plotting $m_{bl}$ (right panel of Fig~\ref{channel1_parton}), which has a kinematic endpoint at $m_{t_2}$.

The presence of such kinematic edges is unique to cascade topologies
of the form in Fig.~\ref{tbw}. This feature has been used extensively in SUSY searches ~\cite{Lester:2006cf}
for heavy neutralino $\chi_2^0$ (equivalent to $H_{1}$ in Fig.~\ref{tbw}), 
which decays into a di-lepton pair (equivalent to pair of quarks in Fig.~\ref{tbw}) and missing energy $\chi_1^0$ (equivalent to X).
Unlike SUSY, however, 
the known masses of the final state particles increases the utility of this variable to a
far greater effect.The combinations of the invariant mass in this channel are bereft of combinatorial uncertainties that are typical in SUSY and other channels discussed below.

% There exists a very minute possibility of $m_{tt}$ edge being possible in MSSM due to the following topology:$\tilde g\tilde g\rightarrow tttth$. Here the $\tilde t_1$ and $t$ are nearly degenerate resulting in extremely light neutralinos. However, this is beset with a combinatorial uncertainty which will significantly smear the presence of an edge.
% In addition, the presence of a complimentary edge in $m_{bl}$ discounts this possibility as sbottoms degenerate with bottom mass is not possible. 

\subsection{Channel 2 : $H_1\rightarrow tt_2 \rightarrow t t Z$} We consider leptonic decay of $Z$-boson while both the tops decay hadronically. Similar to the Channel 1, this mode also exhibits the following features:
\begin{itemize}
 \item \textbf{Kinematic edge in the $m_{tt}$ distribution:} The distribution of the invariant mass of the top pairs has an edge given by \footnote{The distribution of the invariant mass of the top quark which is the
 	daughter of $t_2$ and one of the leptons
 	will also have the kinematic edge similar to $m_{lb}$. However, unlike $m_{lb}$ the identity of top is uncertain and leads to a combinatorial uncertainty. Additionally, the transverse mass of top quark and Z-boson defined by $m_T=\sqrt{m_Z^2+m_t^2+2\left(E^t_TE^Z_T-\textbf{p}^t_T.\textbf{p}^Z_T\right)}$ also has an edge at $m_{t_2}$.}.
 \begin{equation}
\left(m^{edge}_{tt}\right)^2~=~2 m_t^2~+~2\left(E_{t_{a}}E_{t_{b}} + |\textbf{P}_{t_{a}}||\textbf{P}_{t_{b}}| \right)
\label{edgett}
\end{equation}

The magnitude of the transverse momenta of the top quarks in the rest frame of $t_2$ is given by
\begin{eqnarray}
p_{t_a}^2=\frac{m_t^4+m_{t_2}^4+m_{H_1}^4-2\left(m_t^2m_{t_2}^2+m_t^2m_H^2+m_{t_2}^2m_{H_1}^2\right)}{4m_{t_2}^2}\nonumber\\
p_{t_b}^2=\frac{m_t^4+m_{t_2}^4+m_{h}^4-2\left(m_t^2m_{t_2}^2+m_t^2m_h^2+m_{t_2}^2m_{h}^2\right)}{4m_{t_2}^2}
\end{eqnarray}
where $t_{a,b}$ are the two tops for the event and $E_{i}^2=m_i^2+p_{i}^2$.

\end{itemize}

\subsection{Channel 3 : $H_1\rightarrow tt_2 \rightarrow t t h$}  We consider $h\rightarrow b\bar b$ decay mode of the Higgs as it is the most dominant. The final states is characterized by a pair of top and bottom quarks. We consider
one of the tops to decay semi-leptonically that suppresses multi-jet QCD background. Similar to Channel 2 this topology also exhibits a kinematic edge in $m_{tt}$ distributions as well as the transverse mass $M_T$ of the $t,h$ system.

While the parton-level results are all promising, it is still challenging to observe
the edges at the LHC beneath the SM backgrounds with proper identification of the top,
bottom and Higgs.The rest of the analysis is dedicated in identifying a collider strategy which can closely reproduce the parton-level plots in Fig. \ref{channel1_parton}. \\

\section{Identifying edges at the LHC}
\begin{figure}[!t]
	\begin{center}
		\begin{tabular}{c}
			\includegraphics[width=8.2cm]{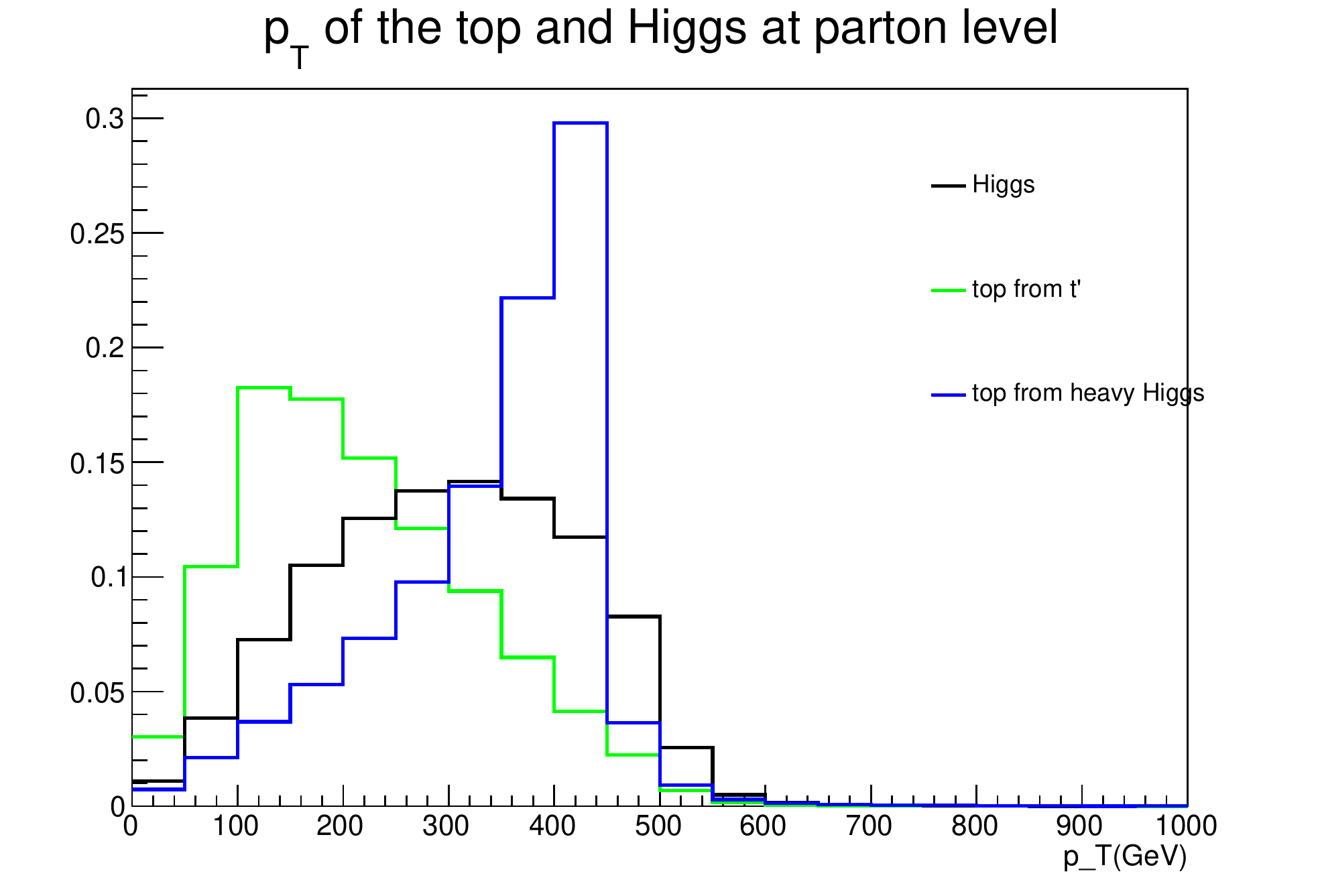}
		\end{tabular}
	\end{center}
	\caption{
		$p_T$ distribution of the top and the Higgs at the parton-level. }
	\label{parton}
\end{figure}

The final state particles in a collider environment are typically identified in terms of leptons, photons, $\tau$ and jets. The strength of the analysis lies, not only in reproducing the parton-level plots presented earlier but also in its effectiveness in reducing the SM backgrounds. In a scenario where top quarks, Higgs, $Z$ and $W$-bosons are boosted, their decay products can be captured inside a cone of radius $R$. The criterion to determine $R$ follows from the fact that the the mass difference between the heavy Higgs $H_{1}$ and the top partner $t_{2}$ must be significantly greater than the top threshold.
i.e
\begin{equation}
\Delta m=m_{H_{1}}-m_{t_{2}}\geq 400~\text{GeV},
\end{equation}
This ensures that the opening angle ($R$) between the top decay products is
\begin{equation} 
R\sim\frac{2m_{t}}{p_T^{t}}\leq 1.5
\label{radius}
\end{equation}
The specific choice of our benchmark points~\ref{benchmark} warrants us such opening angle. 
Fig \ref{parton} shows the $p_T$ distribution of the top quarks and the Higgs for Channel 3 with the second benchmark point. The $p_{T}$ distribution has a peak at about 350 GeV for the leading top and the Higgs. As a result, the top jet, satisfies the condition in Eq.\ref{radius}. The Higgs, on account of its lighter mass will also satisfy the criterion in Eq. \ref{radius}. The slightly larger choice of $R$ ensures that the constituents of the sub-leading top (for the second and third channel) can be captured inside a jet as well.\\

\subsection{Jet Reconstruction:}
The parton-level events for our signal topology are generated with {\tt{MADGRAPH}} at 14 TeV centre of mass energy using PDF {\tt{NNLO1}} \cite{Ball:2012cx}
The events are showered and hadronized using {\tt{PYTHIA}}~\cite{Sjostrand:2007gs}. 
The showered events are then subsequently passed through the {\tt{DELPHES}} detector simulator \cite{deFavereau:2013fsa} using the CMS card.
We extract the calorimetric four vectors for each event using the following acceptance criteria:
\begin{equation}
E_{e-cal}>0.1 ~\text{GeV}\;\;\;;\;\;\;E_{h-cal}>0.5 ~\text{GeV}
\end{equation}
These calorimetric outputs are clustered using {\tt{FASTJET}}~\cite{Cacciari:2011ma}  
with the $Cambridge-Achen$ algorithm~\cite{Dokshitzer:1997in, Bentvelsen:1998ug} to reconstruct fat-jets.
The top candidates in the event are identified using substructures of the reconstructed fat-jets with the jet reconstruction parameter to be $R=1.5$. On account of the large transverse-momentum ($p_T$)
associated with each event, we require 
the jet to have a minimum $p_T$ of 50 GeV. The reconstructed `fat' jets are required to have rapidity in the range $[-2.5,2.5]$. 

All the Channels discussed above are characterized by the presence of at least one top \footnote{In Channels 2 and 3 with two tops, only the leading top is tagged using the top tagger.}. 
There has been tremendous improvement in the techniques of identifying boosted tops using TopTagger~\cite{Plehn:2009rk, Kaplan:2008ie}. We briefly outline the algorithm adapted by us for tagging the top jet;
\begin{itemize}
	\item Scanning through the three leading jet in each event, we identify the top candidate  using {\tt{HEPTOPTAGGER}}~\cite{Plehn:2009rk}. 
	\item The fat-jets passing through the tagger are subject to
	filtering procedure where the constituents of each jet are reclustered with $R_{filt}=0.3$. 
	\item Out of all the subjets inside a jet, only $5 (n_{filt})$ of the hardest subjets are retained.
	\item The invariant mass of the three subjets are required to lie between 150 GeV to 200 GeV.
\end{itemize}

\begin{figure}[!t]
	\begin{center}
		\begin{tabular}{cc}
			\includegraphics[width=8.2cm]{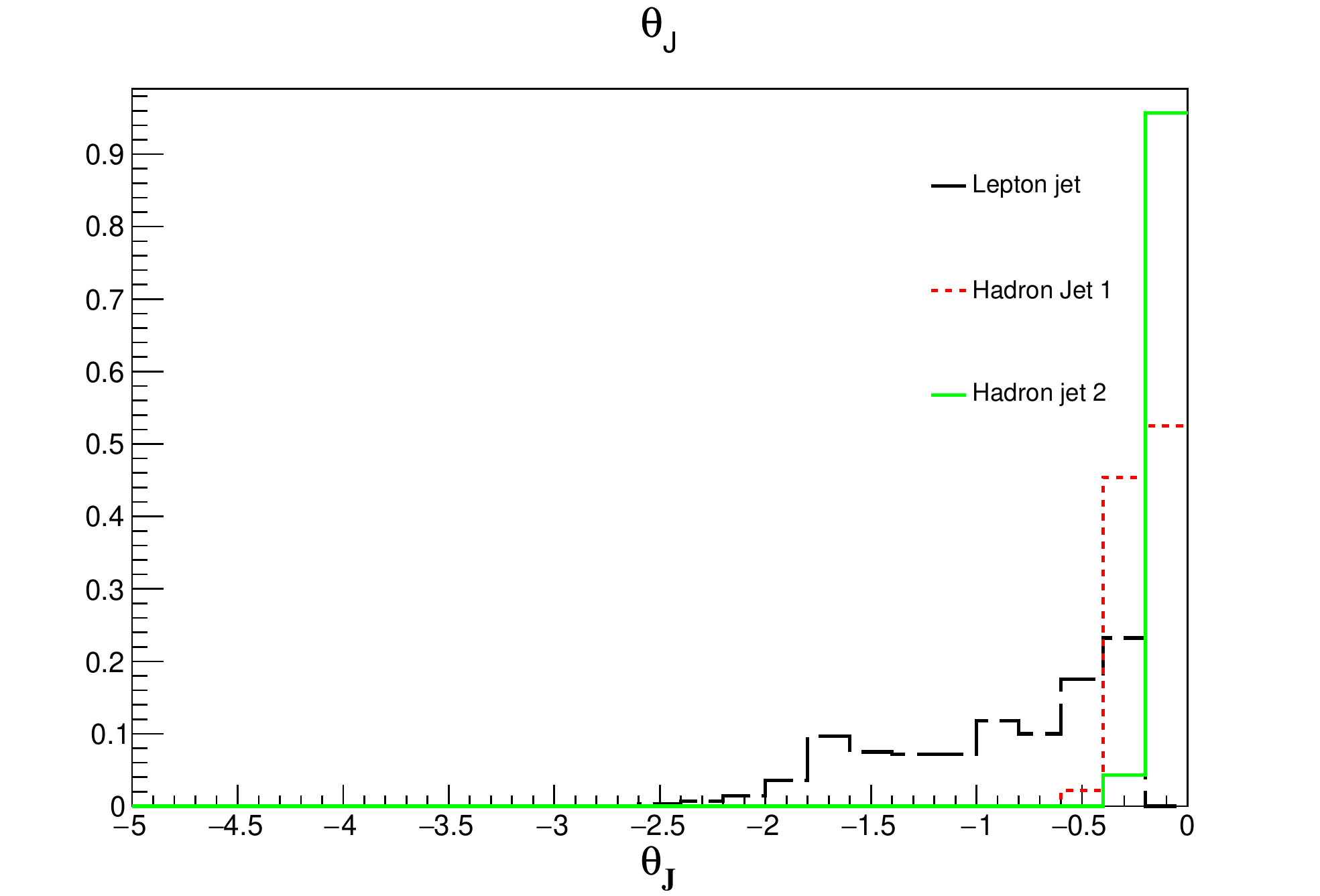}&\includegraphics[width=8.2cm]{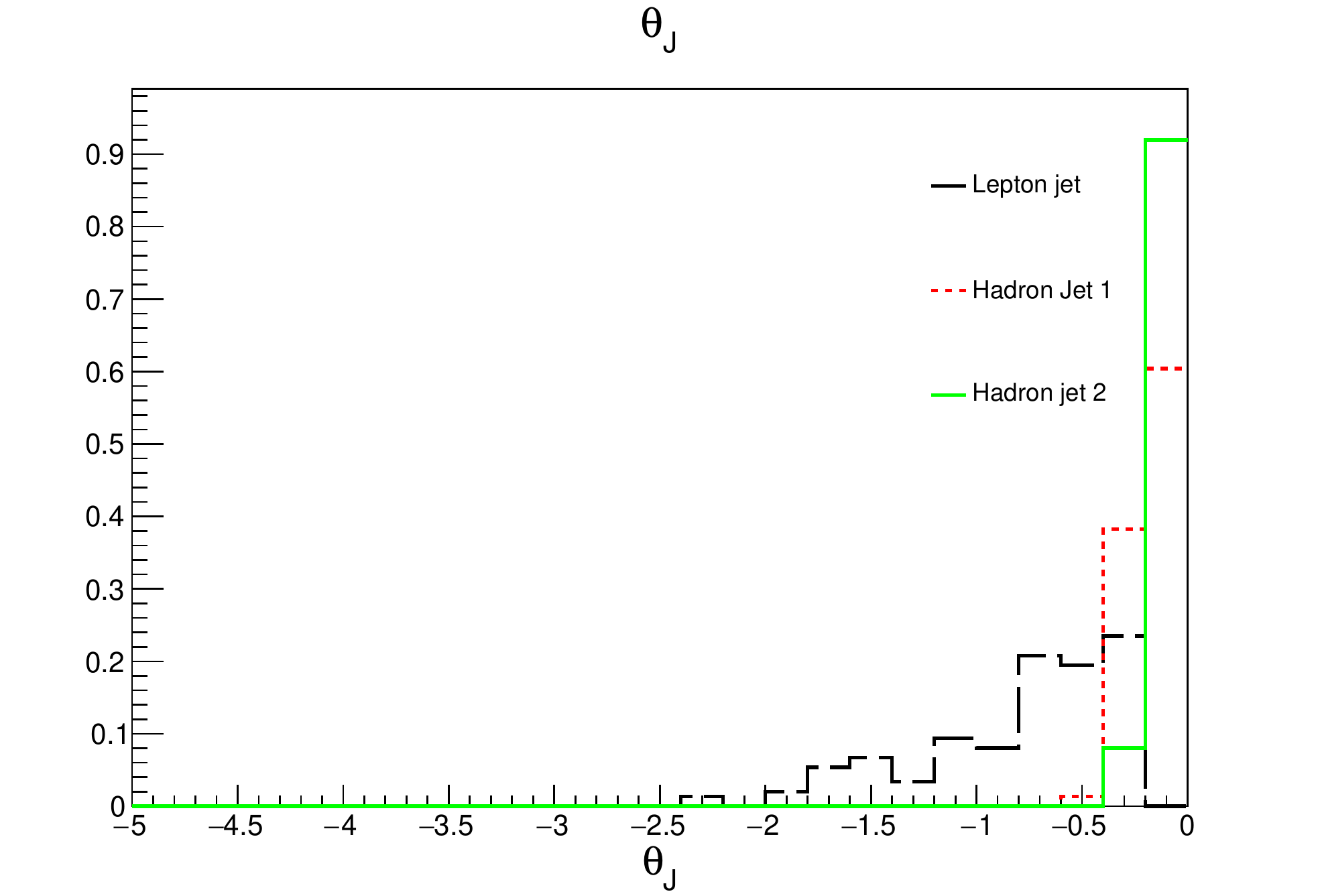}
		\end{tabular}
	\end{center}
	\caption{\it{$\theta_J$ distribution for the three leading jets for Channel 1 (left) and Channel 2 (right)}}
	\label{thetaJ}
\end{figure}

We now discuss the individual strategy adapted for each of the decay channel of the $t_2$:

$\bullet$ $p p~\rightarrow~H~\rightarrow~t t_2,~t_2 \rightarrow b W:$ 
The three leading jet in the event correspond to one of the top, $b-$ jet and the $W-$ jet. 
Only events with a single isolated lepton associated with the decay of W-boson are selected. 
The leptons are isolated with respect to the fat-jets. For each event with a single lepton(at the parton-level), we construct a cone of $\Delta R=0.3$ around the lepton. The leptons are considered to be isolated if the  total energy deposit
within this cone is less than $10\%$ of the transverse momentum of the lepton.
Since we assume the $W$ to decay leptonically, the $W-$ jet can be easily differentiated from the other two by computing the hadronic energy fraction inside the jet defined as
\begin{equation}
\theta_J=\frac{1}{E_j^{total}}\sum_iE_i^{h-cal}
\end{equation}
$E_j^{total}$ is the total energy of the $j^{th}$ jet and $E_i$ is the energy deposited in the $i^{th}$ $h-cal$ cell by a constituent of the $j^{th}$ jet. A $W-$ like jet is likely to deposit all its energy in the electromagnetic calorimeter,  $\theta_J$ is likely to be close to zero. On the other hand, top-like and $b-$ jets deposit most of their energy in the hadronic calorimeter (since we consider hadronic decay of top). This leads to larger values of $\theta_J$ for them. It is convenient to take the logarithm of  $\theta_J$ which further accentuates the difference between jet with or without hadronic activity. 
We identify W-like jet as the fat-jet having minimum hadronic activity. The other two jets are top-like jet and
b-like jet. 
% The three leading jets for the event are then sorted in ascending order of $Log(\theta_J)$, where the jet with the lowest $\theta_J$ is accepted as $W-$ like jet while the other two accepted as top-like jets.
The distribution plotted in the left panel of Fig.~\ref{thetaJ} shows the comparison of $Log(\theta_J)$ of $W-$ like jet and the other hadronic jets (labeled as Hadron Jet 1(2)) for Channel 1.

 As expected, $Log(\theta_J)$ for top-like and $b-$ jets peaks close to zero, while those for $W-$ like jet are large and negative in comparison.
The presence of hadronic activity inside a $W-$ like jet can be attributed to the fact that these are fat ungroomed jets and are likely to collect stray QCD activity.
After the identification of the $W-$ jet we identify the top from the remaining jets through top-tagger discussed above, while the remaining jet is considered to be the $b-$ jet.
We find that a cut of $Log(\theta_J)<-0.3$ on the jet identified as the $W-$like is useful in suppressing
the $t\bar{t}$ + jets background. 

\begin{figure}[!t]
	\begin{center}
		\begin{tabular}{cc}
			\includegraphics[width=8.2cm]{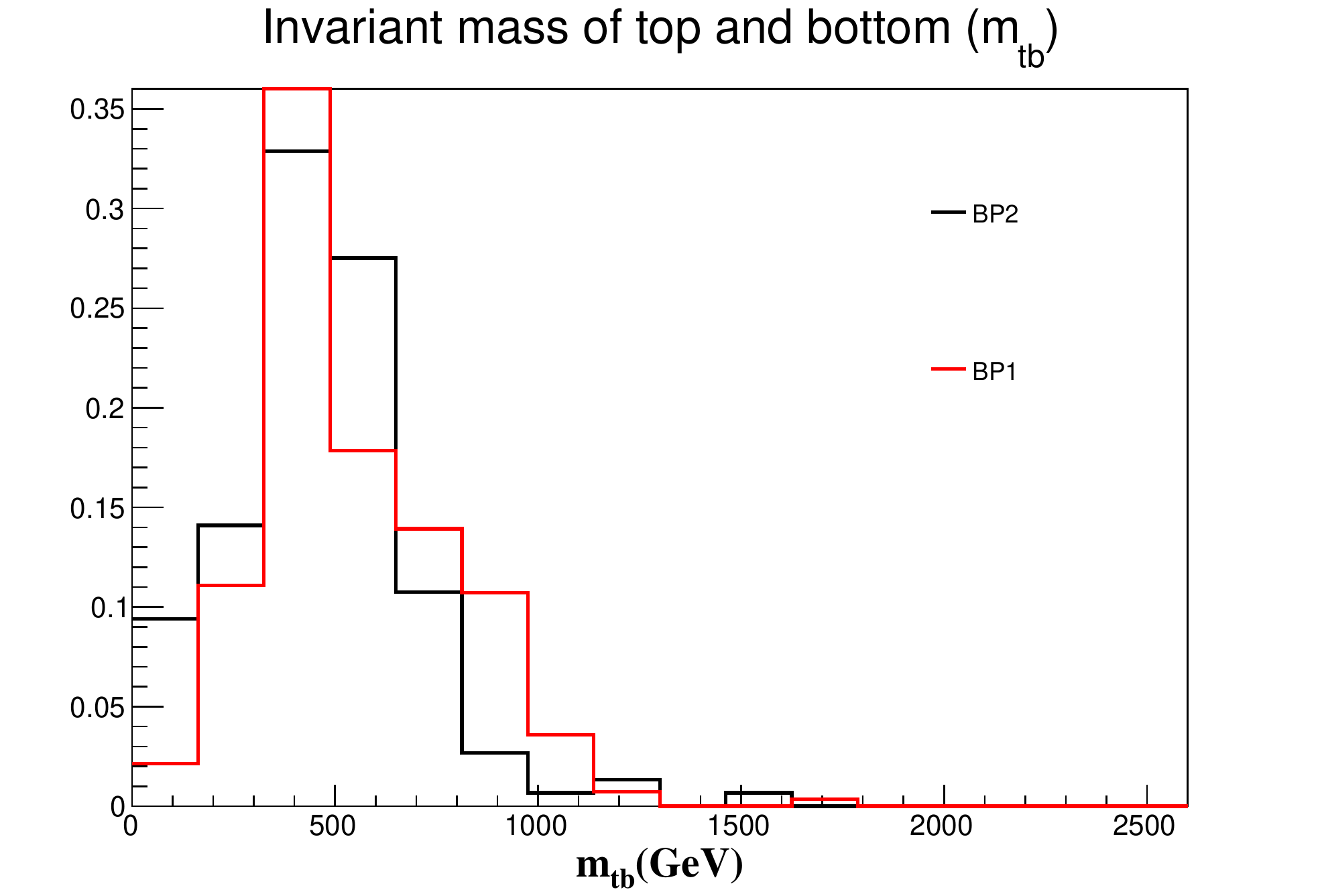}&\includegraphics[width=8.2cm]{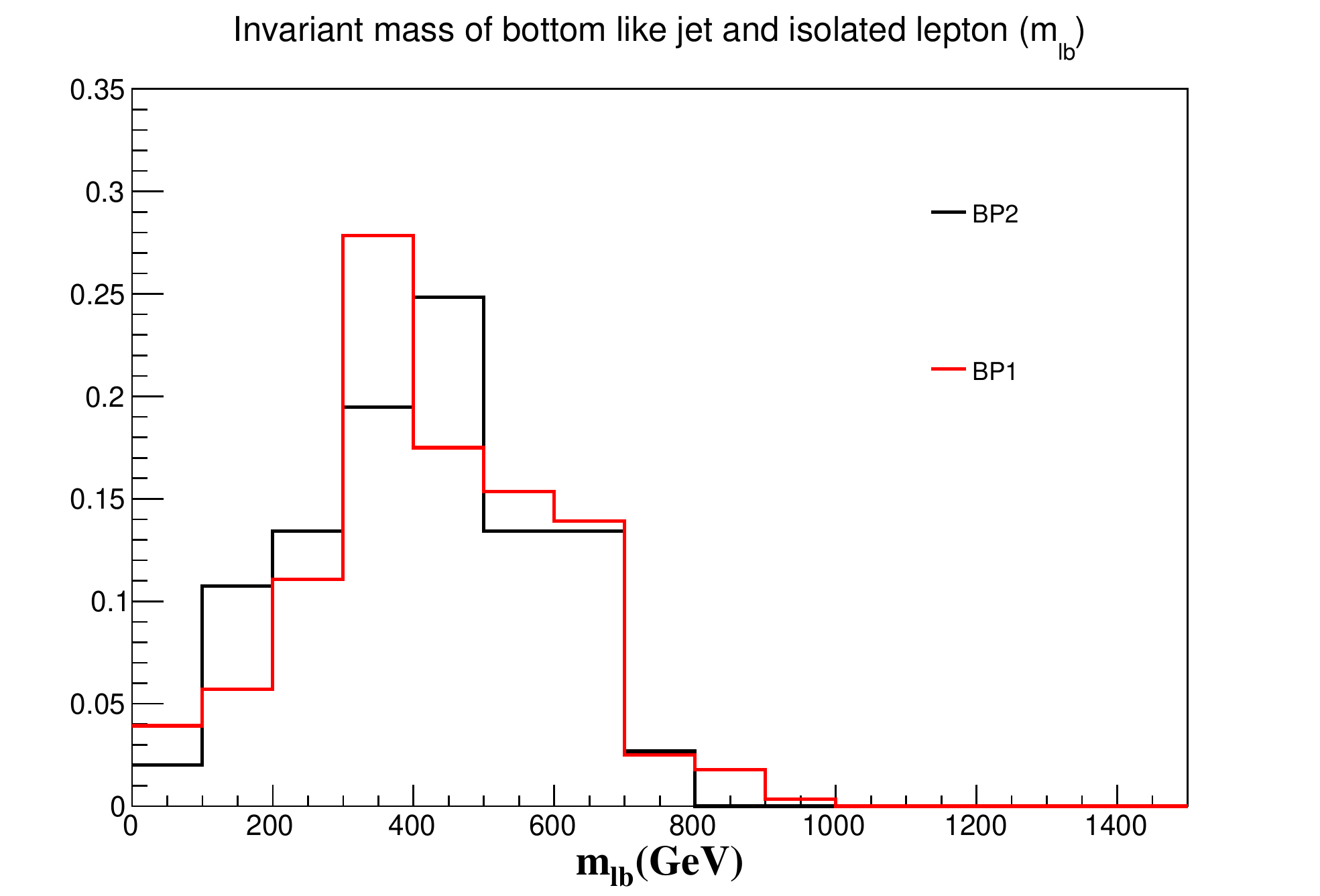}
		\end{tabular}
	\end{center}
	\caption{\it{The simulation plots for $m_{tb}$ (left) and $m_{bl}$ (right) for the two benchmark points.}}  
	\label{channel1}
\end{figure}

The distribution of the invariant mass of the lepton along with the jet not tagged as the top is plotted in the right panel of Fig. \ref{channel1}. 
 Using the edge of $m_{lb}$, we can determine the mass of $t_2$. The distribution of the invariant mass of the two completely hadronic jets is given in the right panel of
Fig.\ref{channel1}. Both these distributions are plotted with 150 signal points. Similar distributions can be obtained for much lesser signal points.
The position of the edge using monte-carlo simulation closely replicates that obtained using the parton-level information, thus highlighting the strength of our analysis.\newline

\begin{figure}[!t]
	\begin{center}
		\begin{tabular}{cc}
		\includegraphics[width=8.2cm]{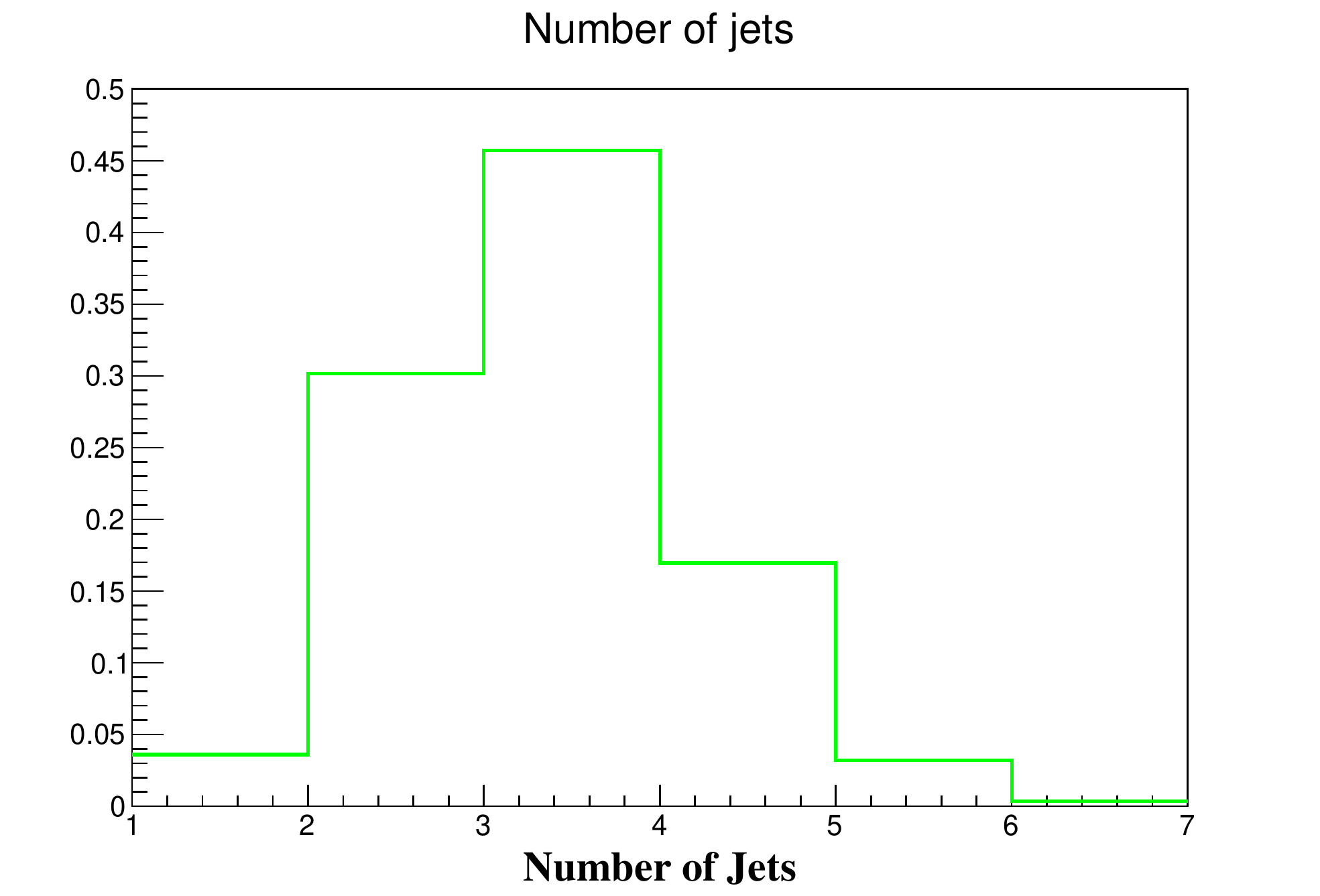}&\includegraphics[width=8.2cm]{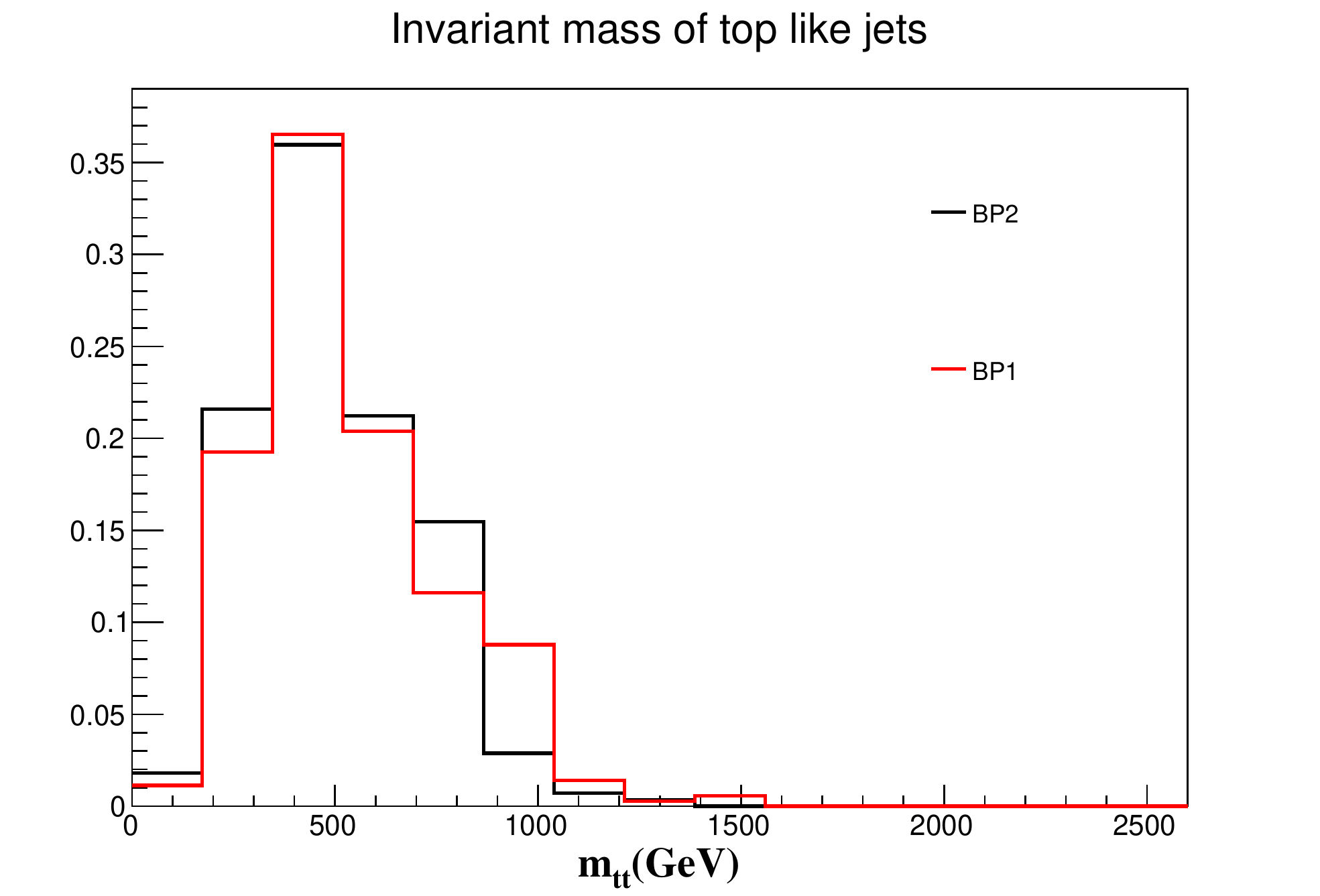}
		\end{tabular}
	\end{center}
	\caption{Left panel shows the distribution of fat-jet multiplicity for channel 2 with BP1. Right panel corresponds to distribution of $m_{tt}$ for this channel}
	\label{jetdist}
\end{figure}

$\bullet$ $p p~\rightarrow~H~\rightarrow~t t_2,~t_2 \rightarrow t Z:$ This channel is characterized by the presence of two top quarks which decay hadronically along with the presence of a $Z$ boson which is assumed to decay leptonically.
The event is triggered by the presence of two isolated leptons. The $Z$ jet can be distinguished from the top jet by computing the hadronic energy fraction inside the jet described above. The right panel of Fig.~\ref{thetaJ} shows the comparison of $Log(\theta_J)$ of $Z-$ like jet and the other hadronic jets (labeled as Hadron Jet 1(2)) for Channel 2. Like earlier, we give a cut of $Log(\theta_J)<-0.3$ on the $Z-$jet.
 This is followed by tagging one of the top out of the two remaining jets.
 The left panel of Fig.~\ref{jetdist} shows the distribution of fat-jet multiplicity.
 Right panel of Fig.~\ref{jetdist} represents the distribution of $m_{tt}$ constructed out of the filtered hadronic jets (with relatively larger hadronic content). The distribution is plotted with 200 signal points. One of the jet is tagged as the top. The plots for both the benchmark points exhibit an edge close to the expected value.\newline

 $\bullet$ $p p~\rightarrow~H~\rightarrow~t t_2,~t_2 \rightarrow b h:$
 The signal is characterized by the presence of two top jets and a Higgs jet. We assume $h\rightarrow b\bar b$ decay mode of the Higgs. Multi-jet QCD background can be suppressed by assuming that one of the top decays leptonically. As a result we select events with a single isolated lepton. Once a top is tagged, we scan over the other leading jets to tag the Higgs using the {\tt{MASSDROP}}\cite{Butterworth:2008iy} tagger outlined below:
 \begin{itemize}
 	\item For a given candidate fat-jet $j$, the last stage of clustering is undone and broken into two subjets $j_1$ and $j_2$ such that $m_{j_1}>m_{j_2}$.
 	\item  In the event of a significant mass drop, $m_{j_1}<\mu m_j$, with a not too asymmetric splitting, $y=\frac{min(p^2_{t_{j_1}},p^2_{t_{j_2}})\Delta R^2_{j_1j_2}}{m_j^2}>y_{cut}$,
 	the jet $j$ is considered to be tagged.
 	
 	\item If the second condition is not satisfied, redefine $j$ to be the subjet $j_1$ and first step is repeated. 
 \end{itemize}

 $\mu$ and $y_{cut}$ are real numbers and are chosen to be $\mu=0.67$ and $y_{cut}=0.09$.
 We retain only those 'Higgs-like jet' whose invariant mass lie within the window of 10 GeV centered about 125 GeV.
 The other top can be reconstructed by assuming neutrinos to be the only source of missing energy for the event. We extract transverse components of the neutrino momentum as the negative of the vector sum of the transverse momentum of 
 all visible particles in an event.
 The $z$-component of the neutrino momentum is extracted by solving the equation for the $W$-boson mass $m_w^2=(p_l+p_\nu)^2$ and is given as
 \begin{equation}
 p_{\nu z}=\frac{1}{2p_{eT}^2}\left(Dp_{eL}\pm E_e\sqrt{D^2-4p_{eT}^2\slashed{E}_T^2}     \right)
 \end{equation}
 where $D=m_w^2+2\bar p_{eT}. \slashed{\bar E}_T $ and we assume $D^2-4p_{eT}^2\slashed{E}_T^2 >0$.
 Once the $z$-component of the `neutrino' momentum is identified, we reconstruct $W$ using the momenta of the isolated lepton and neutrino.
 We identify the anti-kt \cite{Cacciari:2008gp} b-tagged jet (reconstructed with $R=0.5$ and $p_T^{min}=50$ GeV) coming from the second top by demanding that $\Delta R$ between the Higgs-like jet (top-tagged jet) and the b-tagged jet is greater than 1.5.
 Using the b-tagged jet and $W$, we further reconstruct the semi leptonic top.
 
\begin{figure}[!t]
 	\begin{center}
 		\begin{tabular}{cc}
 			\includegraphics[width=8.2cm]{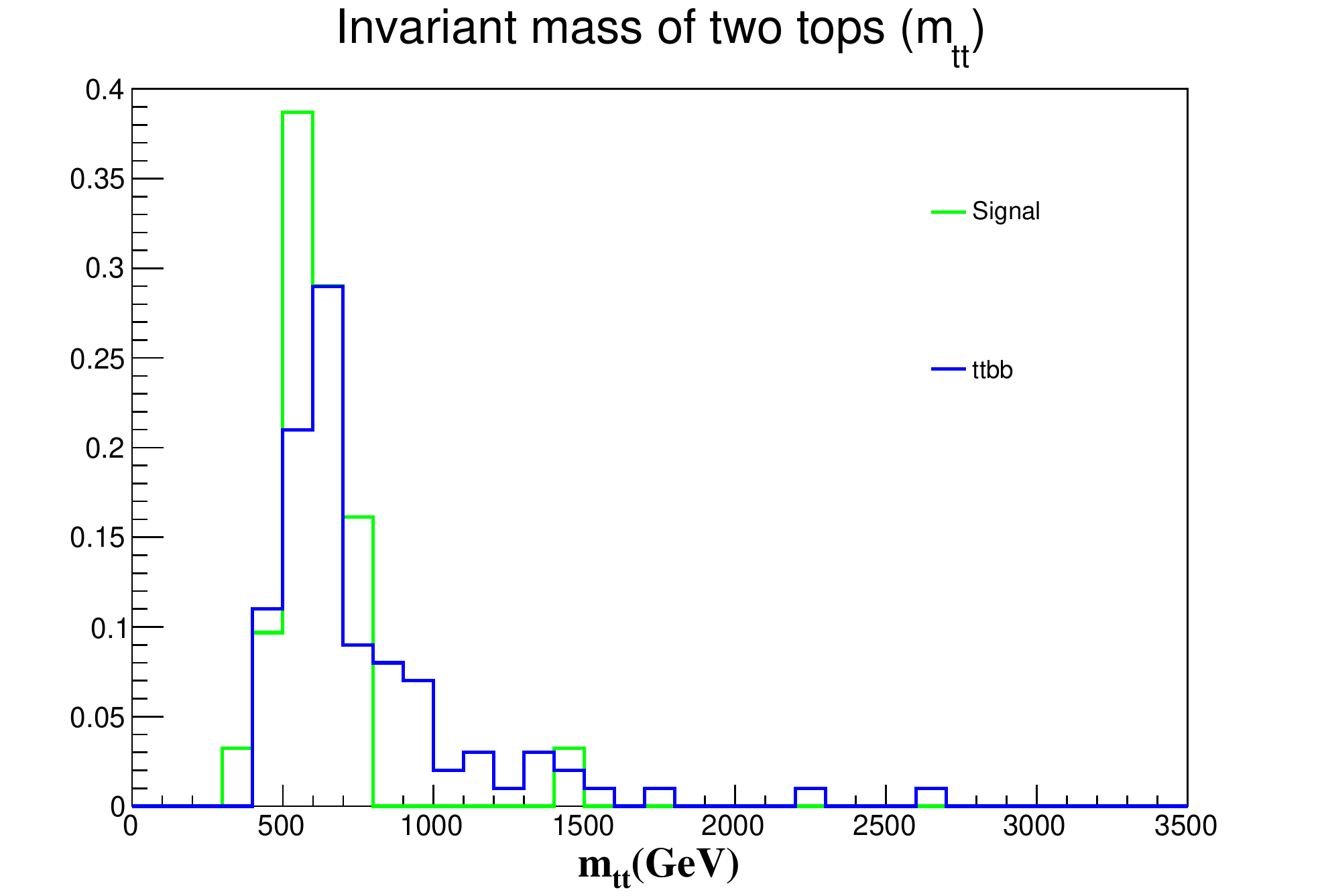}
 			&\includegraphics[width=8.2cm]{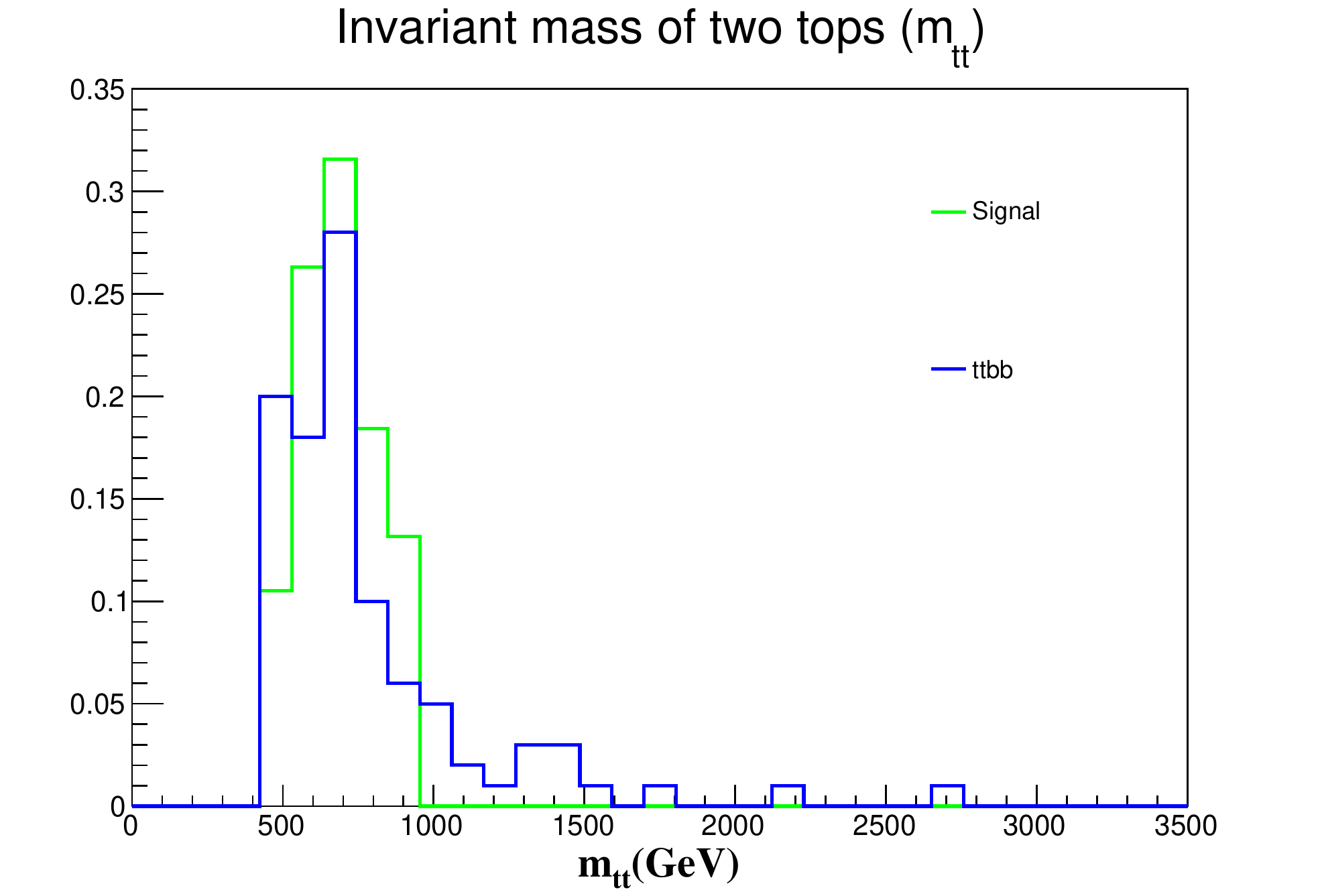}
 		\end{tabular}
 	\end{center}
 	\caption{  $m_{tt}$ distribution for signal (green) and $ttbb$ background (blue) for BP1 (right) and BP2 (left).}
 	\label{edge_lepton}
 \end{figure}

 Fig \ref{edge_lepton} gives the $m_{tt}$ distribution (in green) for the signal with the $ttbb$ background superimposed (in blue).
 Both plots are plotted with about 35 signal points.
 The left plot shows a very distinct edge at $\sim 800$ GeV while the expected edge for $(m_H,m_{t_2})=(1100,700)$ is $839$ GeV. 
 Similar agreement is obtained for the mass combination $(m_H,m_{t_2})=(1200,600)$ where an edge-like feature is seen at $~970$ GeV while the parton-level result is at $1023$ GeV. Thus our simulation can predict the location of the edges to within $10 \%$ of the actual value and can serve as a smoking gun for the existence of such topologies

The final state for this channel is exactly similar to the $t\bar{t}h$ process in the SM
In the event of an observation of the latter, it is an irreducible background for signal topology considered in Channel 3. However, the construction of the $m_{tt}$ invariant mass in the SM would not exhibit a edge like feature like in the case of our signal and hence can be easily distinguished.
In addition we would like to point out that some of the techniques introduced in this work could be beneficial to probe the $t\bar{t}h$ in the SM especially with a boosted Higgs decaying as $h\rightarrow\gamma\gamma$. The `Higgs jet' could be identified with as the one with low $Log(\theta_J)$ similar to the jets with low hadronic content jets in Channel 1 and Channel 2.

 \section{Results and Discussion:}
 \label{results}
 The observation of these distributions require certain number of signal events. In accordance with the branching fractions, we demand a minimum of $\sim$ 50,30,30 signal points for Channel 1,2 and 3 respectively. Table \ref{summary} 
 gives the projected luminosities for the accumulation of these signal points for the calculated acceptance from our simulation. In computing the projected luminosities we assume 60$\%$ branching fraction of $H_1$ into $tt_2$.
 It also gives the predicted and the observed values of the edges for all three channels.
 \begin{table}[h]
 	\renewcommand{\arraystretch}{1.5}
 	\begin{tabular}{cc|c |c |c |c |c|c|c|}
 		\cline{2-9}
 		&\multicolumn{4}{|c}{BP1( 92 $fb$)}&\multicolumn{4}{|c|}{BP2(129 $fb$)}\\
 		\cline{1-9}
 		\hline
 		\multicolumn{1}{ |c|  }{Channel}&Edge$^{obs}$&Edge$^{exp.}$&Efficiency&Luminosity($fb^{-1}$)&Edge$^{obs}$&Edge$^{expec}$&Efficiency& Luminosity($fb^{-1}$)\\ 
 		\hline
 		\multicolumn{1}{ |c|  }{Channel 1($m_{tb}$)}&$\sim$1000&1025  & 0.005&1100 & $\sim$800&830&0.003&1300\\
 		\hline
 		\multicolumn{1}{ |c|  }{Channel 2($m_{tt}$)}&$\sim$1000& 1036 & 0.007&$>3000$ & $\sim$850&847&0.005&3000\\
 		\hline
 		\multicolumn{1}{ |c|  }{Channel 3($m_{tt}$)}&$\sim$970&  1025& 0.0008&$>3000$& $\sim$800&840&0.0008&3000\\
 		\hline
 	\end{tabular}
 	\caption{Reaches and predictions of the edges for three different channels. The cross-sections in brackets correspond to the gluon-gluon fusion production rate of the heavy resonance for the benchmark points at 13 TeV $N^3$LO.} 
 	\label{summary}
 \end{table}
 
The smaller efficiency for Channel 3 can be attributed to the fact that in addition to the top tagging we also require the Higgs jet to be tagged. In addition the leptonic top is reconstructed from the $b$ tagged jet which necessarily must not lie inside either of the top tagged or the Higgs tagged jet. The efficiencies for the first two channels are on the lower side due to a cut on $\theta_J$. Higher efficiencies can be obtained by either relaxing or completely ignoring the cut. However, this cut is highly essential in suppressing the $t\bar t+jets$ background which may possibly smear the edges.
 Channel 1 offers the most optimistic scenario for both the benchmark points observables at High Luminosity (HL) phase of the LHC. Additionally, this channel is free from any combinatorial uncertainties for the construction of the second edge ($m_{lb}$) which complements the $m_{tb}$ distribution. Both these aspects make it an exciting prospect to explore.
 
\begin{figure}[!t]
	\begin{center}
		\begin{tabular}{cc}
			\includegraphics[width=8.2cm]{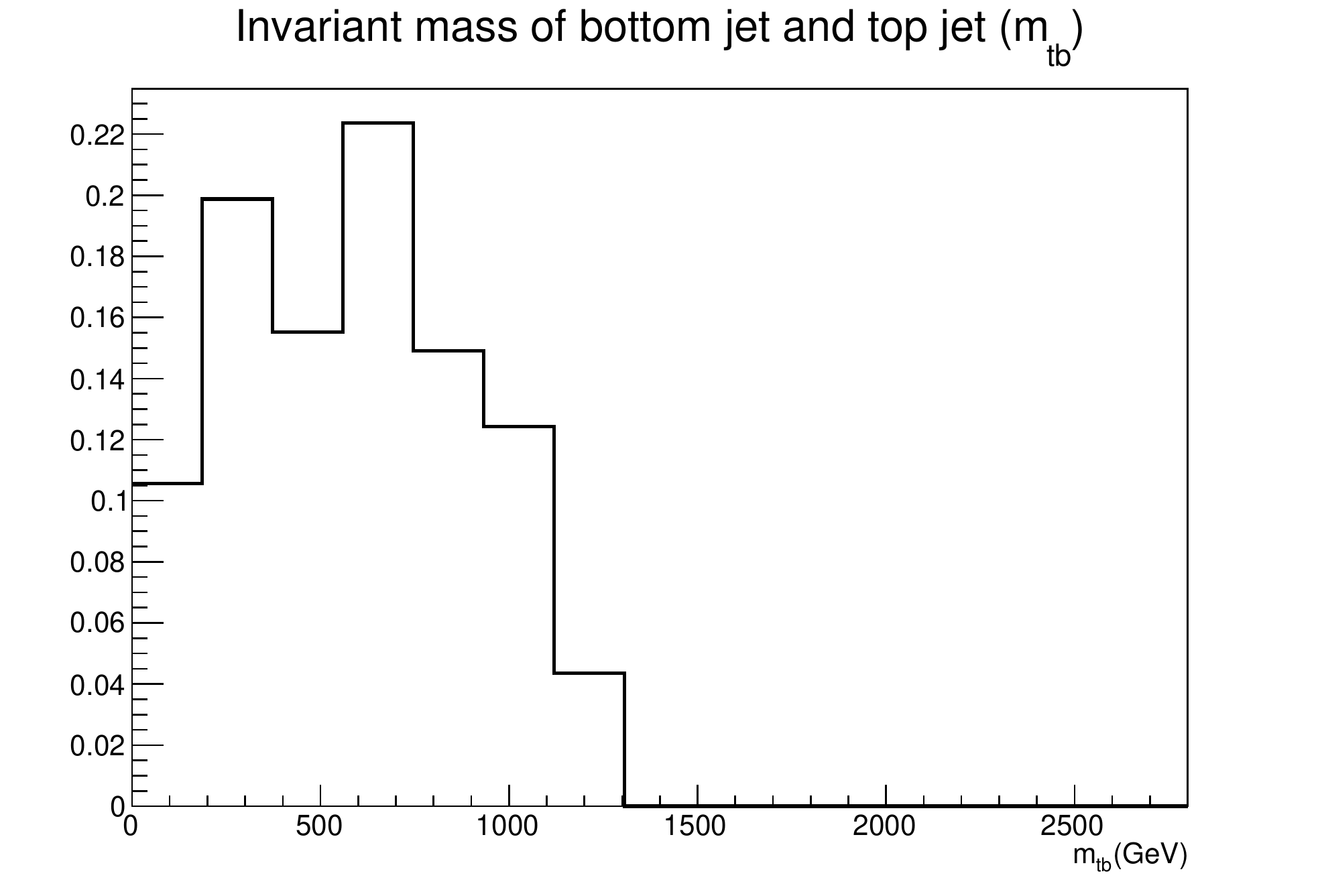}&\includegraphics[width=8.2cm]{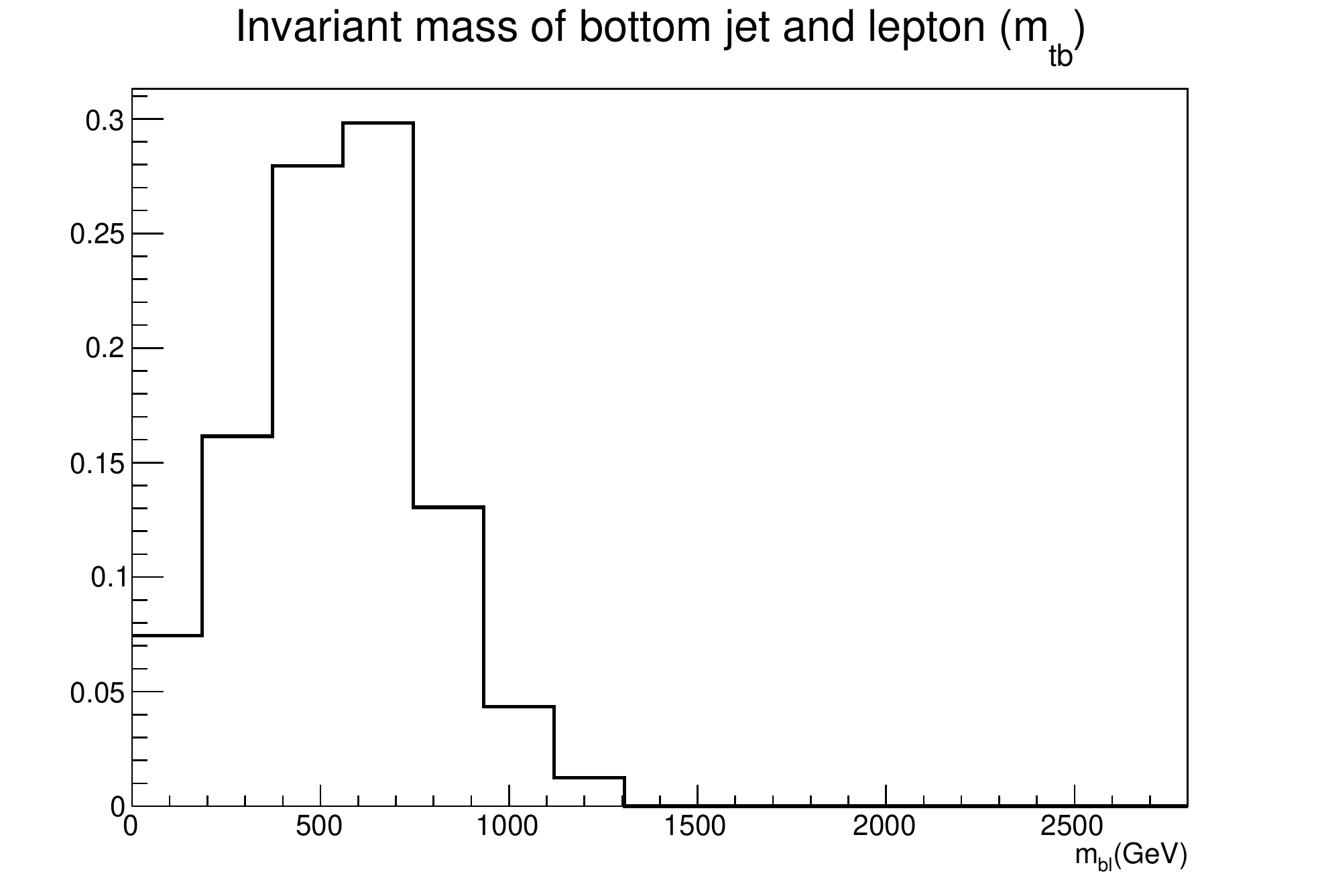}
		\end{tabular}
	\end{center}
	\caption{\it{The simulation plots for $m_{tb}$ (left) and $m_{bl}$ (right) for the third benchmark point $(m_{H_1},m_{t_2})=(1500,1000)$ GeV.}}  
	\label{edge_bp3}
\end{figure}

 The analysis from Run-I of the LHC constrain the masses of the top partner to be $>950$ GeV \cite{Aad:2016qpo}.
 In light of this, we implement a scenario
 $(m_{H_1},m_{t_2})=(1500,1000)$ GeV. Due to the small production cross section of heavy scalar ($37 fb^{-1}$) for this mass, only the leading decay mode of $t_2\rightarrow Wb$ is relevant in this case. The edges for this bench mark point in given in Fig. \ref{edge_bp3} and Table \ref{summary1} gives a summary of results for the same.
 
  \begin{table}[h]
  	\renewcommand{\arraystretch}{1.5}
  	\begin{tabular}{|c|c|c|c |c|c|}
  		\cline{3-6}
  	\multicolumn{2}{c}{}	&\multicolumn{2}{|c}{$m_{tb}$}&\multicolumn{2}{|c|}{$m_{tl}$}\\
  		\cline{1-6}
  		\hline
  		Efficiency&Luminosity($fb^{-1}$)&Edge$^{obs}$&Edge$^{exp.}$&Edge$^{obs}$&Edge$^{expec}$\\ 
  		\hline
  		0.0044&2500&1100&1103&900&997\\
  		\hline
  		\end{tabular}
  	\caption{Table shows the observed and the expected value of the edge for $m_{tb}$ and $m_{bl}$ invariant mass distributions for the third benchmark point. The luminosity corresponds to the accumulation of 40 signal points.
  		The cross-sections in brackets correspond to the gluon-gluon fusion production rate of the heavy resonance for the benchmark points at 13 TeV $N^3$LO.} 
  	\label{summary1}
  \end{table}
  
 The expected reach to accumulate $40$ signal points is about $2.5~ab^{-1}$. 
  It is interesting to stress at this point that this technique is not restricted to the case with heavy scalars. The analysis can be repeated with more massive colored objects (KK excitations of gluons), which enjoy two advantages:\\
 a) Colored objects enjoy large production cross section at heavier masses.  A heavy scalar of mass 1.4  TeV  has cross section similar to a 3 TeV KK gluon. The edges corresponding to this mass will be at much heavier scales resulting in much less smearing.\\
 b) It also increases the sensitivity to probe much heavier masses for the vector like top partner, much beyond the limit possible by the high luminosity LHC.
 This opens up a lot of interesting possibilities and will be addressed in an upcoming publication \cite{KKg}.

  It is important to note that at this stage the cuts are not tuned to get the desired $S/\sqrt{B}\sim 5$ for the leptonic case.
  They are fashioned to get the desired kinematic distributions with enough signal points at much lower luminosities. Observation of these distributions would serve as a smoking gun to tighten the selection around the probable masses to achieve the desired significance.
 
 We find that the analysis discussed thus far serves to achieve a multi-fold objective:
 a) Edges are typically constructed out of leptonic final states which have  a sharp feature owing the distinct determination of the lepton momenta . In this work we have constructed edges out of top and bottom jets which are likely to exhibit  smearing, even for the signal. In this work we have successfully demonstrated the construction of these edges using jets and achieved a fair degree of success to this effect. The quality of the edges can be improved further by imposing $b-$ tagging criteria. 
 b) A definite pointer towards the existence of new physics scenarios. 
 This, can be further extended to argue that it is an indicator towards the existence of non-MSSM scenarios. c) Gives a hint towards the region of parameter space where such new physics resonances can be expected to lie. 
 An analysis of this nature has an extremely wide scope in general. Looking out for the existence of new physics by identifying characteristic unique to it can serve as a trigger which may aid the direct searches for the current and future runs.\\
 \textbf{Acknowledgements:} We would like to thank Amit Chakraborty, Monoranjan Guchait and Tuhin Roy for useful discussions. We also thank N. Manglani for collaboration in the preliminary stages of the work. We would also like to thank Department of Theoretical Physics, TIFR for the use of its computational resources.

\bibliography{tth.bib}
\end{document}